\documentclass[twocolumn,showpacs,preprintnumbers,amsmath,amssymb]{revtex4}

\usepackage{graphicx}
\usepackage[dvips]{color}

\begin{document}

\hyphenation{Fesh-bach}


\title{Determination of atomic scattering lengths from
measurements of molecular binding energies near Feshbach resonances}

\author{A. D. Lange$^{1}$, K. Pilch$^{1}$, A. Prantner$^{1}$, F. Ferlaino$^{1}$, B. Engeser$^{1}$, H.-C. N\"{a}gerl$^{1}$, and R. Grimm$^{1,2}$}
\affiliation{$^{1}$Institut f\"{u}r Experimentalphysik und Zentrum
f\"{u}r Quantenphysik,
Universit\"{a}t Innsbruck, Innsbruck, Austria\\
$^{2}$Institut f\"{u}r Quantenoptik und Quanteninformation, \"{O}sterreichische
Akademie der Wissenschaften,  Innsbruck, Austria}
\author{C. Chin}
\affiliation{Department of Physics and James Franck Institute,
University of Chicago, Chicago, Illinois 60637, USA}

\date{\today}

\begin{abstract}

We present an analytic model to calculate the atomic scattering length near a
Feshbach resonance from data on the molecular binding energy. Our approach
considers finite-range square-well potentials and can be applied near broad,
narrow, or even overlapping Feshbach resonances. We test our model on Cs$_2$
Feshbach molecules. We measure the binding energy using magnetic-field
modulation spectroscopy in a range where one broad and two narrow Feshbach
resonances overlap. From the data we accurately determine the Cs atomic
scattering length and  the positions and widths of two particular resonances.

\end{abstract}

\pacs{03.75.Hh, 05.30.Jp, 34.50.-s, 21.10.Dr}


\maketitle\narrowtext\section{Introduction}

In experiments on ultracold quantum gases, control of the atomic scattering
length near Feshbach resonances has become a powerful tool to explore different
interaction regimes \cite{FBreview}. Prominent examples include the implosion
and explosion of Bose-Einstein condensates \cite{wieman2001}, the creation of
ultracold molecules \cite{FM,ferlbook}, strongly interacting Fermi gases
\cite{xoverexp,giorgrevi}, and the observation of three-body Efimov states
\cite{Kraemer2006}. Moreover, Feshbach resonances have been proposed as a very
sensitive probe to the variation of fundamental constants \cite{fundamental}.

For all such experiments precise knowledge of the atomic scattering length $a$
is desirable. Usually a multi-channel molecular potential model is employed to
calculate the scattering length. This approach requires experimental input, as
usually provided by photoassociative or Feshbach spectroscopic data
\cite{Tiesinga1996, Abraham1997, Marte2002}. It captures the global scattering
behavior of the system and has a predictive power on Feshbach resonance
positions with an uncertainty of typically a few hundred milligauss. In many
experimental situations, however, precise knowledge of $a$ is needed in the
vicinity of specific Feshbach resonances at a level not reachable with this
global approach.

We here consider the situation in which the scattering length $a$ near a
specific resonance is calculated using molecular binding energy data. Nowadays
powerful experimental methods are available to measure the binding energy of
Feshbach molecules with very high accuracy, such as radio-frequency and
microwave spectroscopy \cite{regal2003, boundbound,
Claussen2003,Thompson2005,Papp2006}. For the relation  between the binding
energy $E_{\rm b}$ and $a$ a simple analytic form can be given in the universal
limit of very large positive values of $a$. It is given by $E_{\rm
b}=\hbar^2/(2 m_r a^2)$, with $m_r$ being the reduced mass. In general, the
relation between molecular binding energy and scattering length is more
complex, depending also on short-range parameters of the molecular potentials.
In principle one may also use a multi-channel calculation to precisely derive
$a$ from measurements of $E_{\rm b}$. When the specific scattering properties
are  needed near a particular Feshbach resonance, a much more simple model can
be applied.

In this Article, we describe a versatile analytic model to directly convert
molecular binding energy data into atomic scattering lengths near Feshbach
resonances. Our approach considers square-well model potentials and can be
applied to any broad or narrow Feshbach resonance, or even to the case of
overlapping resonances. We demonstrate our model on Cs$_2$ Feshbach molecules.
Cesium is an excellent candidate to test scattering models because of its rich
interaction properties \cite{Chin2004, Mark2007}. From fitting the binding
energy data, we show that the magnetic-field dependent scattering length can be
determined with high precision. We compare our results with the full
multi-channel numerical calculation of Ref.\,\cite{Chin2004}.

\section{Feshbach resonance model}

We employ a two-channel square well potential $\hat{V}(R)$ to describe
interacting atoms and weakly-bound molecules near a Feshbach resonance. The
well size is chosen to account for the long range behavior of the molecular
potential, which is dominated by the van der Waals (vdW) interaction
$V(R)=-C_6/R^6$, where $C_6$ is the vdW coefficient and $R$ is the atomic
separation. The associated vdW length scale and vdW energy scales are
$R_{\mathrm{vdW}}=\frac12(2 m_r C_6/\hbar^2)^{1/4}$ and
$E_{\mathrm{vdW}}=\hbar^2/2 m_r R_{\mathrm{vdW}}^2$.

To fully capture the threshold behavior of atomic scattering with vdW
interaction, the well size is chosen to be the mean scattering length
$\bar{a}=4\pi\Gamma(1/4)^{-2}R_{\mathrm{vdW}}$ \cite{Gribakin1993}, where
$\Gamma(x)$ is the gamma function. This choice yields the same binding energy
of weakly-bound molecules in the vdW potential, namely, $E_b=\hbar^2/2 m_r
(a-\bar{a})^2$ for $E_b \ll E_{\mathrm{vdW}}$ \cite{Gribakin1993}.

\begin{figure}
\includegraphics[width=2.8in]{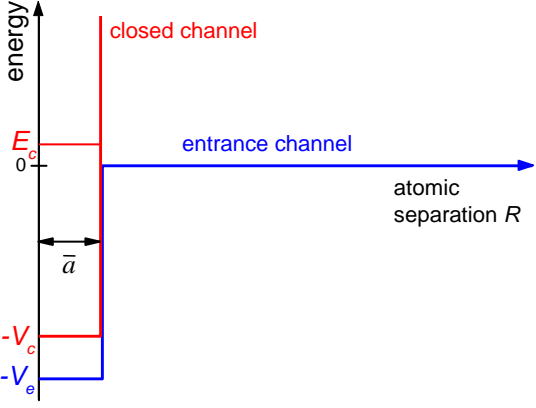}
\caption{(color online) Two-channel square well model of Feshbach
resonances. Scattering states are initially prepared in the entrance
channel $|e\rangle$; the closed channel $|c\rangle$ supports a bare
bound state at energy $E_c$. Feshbach coupling is modeled by the
mixing of the two channels, which is resonantly enhanced when the
bound state is near the scattering state. The interaction range is
set to be the mean-scattering length of the atomic van der Waals
potential, see text.} \label{fig:2chan_fig1}
\end{figure}

In our model, scattering atoms are initially prepared in one spin
configuration, called the ``entrance channel'' $|e\rangle$. A second ``closed
channel'' $|c\rangle$ supports a molecular bound state, see
Fig.\,\ref{fig:2chan_fig1}. The quantum state of an atomic pair with energy $E$
is described as $|\psi\rangle=\psi_c(R)|c\rangle+\psi_e(R)|e\rangle$, which
satisfies Schr\"{o}dinger's equation:

\begin{eqnarray}
\left(-\frac{\hbar^2}{2 m_r}\nabla^2+\hat{V}\right)|\psi\rangle=E|\psi\rangle
\label{2chan_hamiltonian}
\end{eqnarray}

\noindent where, using a matrix representation, we write

\begin{eqnarray}
\hat{V}&=&\left[
\begin{array}{cc}
-V_c & \hbar\Omega \\
\hbar\Omega & -V_e  \end{array} \right]\mbox{ for $R<\bar{a}$,}\nonumber \\
&=&\left[
\begin{array}{cc}
\infty & 0 \\
0 & 0 \end{array} \right]\,\,\,\mbox{ for $R>\bar{a}.$} \nonumber
\end{eqnarray}

Here for $R<\bar{a}$, we assume the attractive potential can support multiple
molecular states, that is, $V_e,V_c\gg E_{\mathrm{vdW}}$, and $\hbar\Omega$
induces Feshbach coupling between the channels. For $R>\bar{a}$, entrance and
closed channel thresholds are set to be $E=0$ and $E=\infty$, respectively. See
Fig.\,\ref{fig:2chan_fig1}.

\subsection{Scattering state above threshold and scattering length}

For $E=\hbar^2k^2/2 m_r>0$, the solution of Eq.\,(\ref{2chan_hamiltonian})
corresponds to an eigenstate  lying above the entrance-channel threshold and is
given by

\begin{eqnarray}
|\psi\rangle&\propto&\frac{\sin (q_+R)}{R}|+\rangle+\frac{A\sin
(q_-R)}{R}|-\rangle
\,\,\,\mbox{for $R<\bar{a}$} \label{2chann_small} \\
&\propto&\frac{\sin (k R+\eta)}{R}|e\rangle \,\,\,\mbox{ for
$R>\bar{a}$}, \label{2chann_large}
\end{eqnarray}

\noindent where $A$ is a constant, $\hbar^2q_{\pm}^2/2
m_r=E+\frac12(V_e+V_c)\pm\frac12(V_e-V_c)\sec2\theta$,
$\tan2\theta=2\hbar\Omega/(V_e-V_c)$, $|+\rangle=\cos\theta
|e\rangle+\sin\theta|c\rangle$ and $|-\rangle=-\sin\theta
|e\rangle+\cos\theta|c\rangle$. The scattering phase shift $\eta$ can be
derived from the continuity condition of the wavefunction at $R=\bar{a}$, which
gives

\begin{equation}
\frac{k}{\tan{(k\bar{a}+\eta)}}=\frac{q_+\cos^2\theta}{\tan{(q_+\bar{a})}}+\frac{q_-\sin^2\theta}{\tan{(q_-\bar{a})}}.
\label{boundary_condition}
\end{equation}

Potential depths $V_c$ and $V_e$ can be chosen such that in the weak
coupling limit $\theta\rightarrow 0$, the scattering phase shift
approaches the off-resonant value $\eta_{\mathrm{bg}}$ and the
closed channel bound state is at $E_c$. These two conditions, given
by

\begin{eqnarray}
\frac{q_+}{\tan{(q_+\bar{a})}}=\frac{k}{\tan{(k\bar{a}+\eta_{\mathrm{bg}})}} \label{summary1} \\
\sin\left(\hbar^{-1} \sqrt{2m_r(E_c+V_c)} \bar{a}\right) =0, \label{summary2}
\end{eqnarray}

\noindent allow us to eliminate $q_+$ and $q_-$ in
Eq.\,(\ref{boundary_condition}). Assuming further that only one bare bound
state is near the continuum $|E_c| \ll \sqrt{V_cE_{\rm vdW}}$, and the Feshbach
coupling is weak $|\theta|\approx \hbar\Omega/|V_e-V_c|\ll 1$, we get

\begin{eqnarray}
\cot(k\bar{a}+\eta)=\cot(k\bar{a}+\eta_{\mathrm{bg}})-\frac{\Gamma/2}{k\bar{a}E_c},
\label{combine}
\end{eqnarray}

\noindent where we have introduced the Feshbach coupling strength
$\Gamma/2=2\theta^2 V_c$.


Scattering lengths can be derived from the threshold relation
$a=-\lim_{k\rightarrow0}(k\cot\eta)^{-1}$. From Eq.\,(\ref{combine}), we get

\begin{eqnarray}
\frac1{a-\bar{a}}=\frac1{a_{\mathrm{bg}}-\bar{a}}+\frac{\Gamma/2}{\bar{a}E_c},
\label{2chan_a}
\end{eqnarray}

\noindent where
$a_{\mathrm{bg}}=-\lim_{k\rightarrow0}(k\cot\eta_{\mathrm{bg}})^{-1}$
is the background scattering length.

Magnetic tunability of Feshbach resonances can be modeled by a
linear Zeeman shift of the bare bound state as $E_c=\delta
\mu(B-B_c)$, where $B$ is the magnetic field, $\delta \mu$ is the
relative magnetic moment of the two channels, and $B_c$ is magnetic
field that tunes the bare state to the entrance channel threshold.
This magnetic field dependence allows us to rewrite the scattering
length in the standard resonance form
$a=a_{\mathrm{bg}}[1-\Delta/(B-B_0)]$. Here the resonance width
$\Delta$ and position $B_0$ are given by

\begin{eqnarray}
\Delta&=&\delta\mu^{-1}\alpha\frac{\Gamma}2, \label{shift_b}\\
B_0&=&B_c-\beta\Delta,  \label{shift_a}
\end{eqnarray}

\noindent where
$\alpha=(a_{\mathrm{bg}}-\bar{a})^2/a_{\mathrm{bg}}\bar{a}$ and
$\beta=a_{\mathrm{bg}}/(a_{\mathrm{bg}}-\bar{a})$ are dimensionless
parameters.

\subsection{Bound molecular state below threshold} We assume that a bound
eigenstate $|\psi_m\rangle$ is located near and below the continuum with
binding energy $E_b=\hbar^2k_m^2/2 m_r$. Solving
Eq.\,(\ref{2chan_hamiltonian}), we obtain the wave function as

\begin{eqnarray} \nonumber
|\psi_m(R>\bar{a})\rangle&\propto& \frac{\exp(-k_mR)}{R}\,|e\rangle, \\
|\psi_m(R<\bar{a})\rangle&\propto& \frac{\sin
(\bar{q}_+R)}{R}|+\rangle+\frac{A_m\sin (\bar{q}_-R)}{R}|-\rangle, \nonumber
\end{eqnarray}

\noindent where $A_m$ is a constant and
$\bar{q}_\pm=(q_\pm^2-k^2-k_m^2)^{1/2}$.

The binding energy $E_b$ can be determined from the boundary conditions at
$R=\bar{a}$. Following the same calculation as that leading to
Eq.\,(\ref{2chan_a}), we find that for large background scattering length
$|a_{\mathrm{bg}}|\gg \bar{a}$, the wave-number $k_m$ determining the binding
energy $E_b=\hbar^2k_m^2/2 m_r$ satisfies

\begin{equation}
k_m=\frac1{a_{\mathrm{bg}}-\bar{a}}+\frac{\Gamma/2}{\bar{a}(E_b+E_c)}.
\label{2chan_em}
\end{equation}

In the limit when the Feshbach coupling approaches zero $\Gamma\rightarrow0$,
the eigenenergies reduce to $E_b=-E_c$ and $E_b=\hbar^2/2
m_r(a_{\mathrm{bg}}-\bar{a})^2$ (for $a_{\mathrm{bg}}>\bar{a}$), which
correspond to the uncoupled bare states in the closed and entrance channels,
respectively.

One can now convert the binding energy into the scattering length with the
following strategy. First, the binding energy data are fitted with
Eq.\,(\ref{2chan_em}). The fitting parameters are $\delta\mu$, $\Gamma$,
$a_{\rm bg}$, and $B_{c}$. Usually the mean scattering length $\bar{a}$ is
fixed because of sufficient knowledge of $C_6$. In some cases, also
$a_{\mathrm{bg}}$ or $\delta\mu$ are known so that they can be kept fixed
during the fit procedure. Second, one determines the width $\Delta$ and the
location of the Feshbach resonance $B_0$  from the fitting parameters using
Eq.\,(\ref{shift_b}) and (\ref{shift_a}).






\subsection{Multiple overlapping Feshbach resonances}

The above discussed model can be generalized to the case of
multiple, overlapping Feshbach resonances, which commonly occurs in
cold collisions of alkali atoms and chromium atoms.

Let us assume that $N$ weakly-bound states are supported by $N$ independent
closed channels. If the mixing between the closed channels and the entrance
channel are weak and $|a_{\mathrm{bg}}|\gg\bar{a}$, we can extend
Eq.\,(\ref{2chan_hamiltonian}) to model the scattering length and molecular
energy. The results are

\begin{eqnarray}
  \frac1{a-\bar{a}}&=&\frac1{a_{\mathrm{bg}}-\bar{a}}+\frac1{\bar{a}}\sum^N_i\frac{\Gamma_i/2}{E_i}
  \label{multi_a} \\
k_m&=&\frac{1}{a_{\mathrm{bg}}-\bar{a}}+\frac1{\bar{a}}\sum^N_i\frac{\Gamma_i/2}{E_b+E_i},
  \label{multi_em}
\end{eqnarray}

\noindent where $E_i$ is the energy of the $i$th bare bound state and
$\Gamma_i$ is its Feshbach coupling strength to the entrance channel. The
results presented here bear similarity to those from the classic Fano theory on
the coupling of discrete states to a continuum \cite{fano1961}.

Assuming the bare states can be linearly tuned magnetically, namely,
$E_i=\delta \mu_i(B-B_{c,i})$, where $\delta \mu_i$ is the relative magnetic
moment and $B_{c,i}$ is the crossing field value of the $i$-th bare bound
state, we can express the scattering length as $a/a_{\mathrm{bg}}=1-P(B)/Q(B)$,
where $P(x)$ and $Q(x)$ are polynomials of $(N-1)$- and $N$-th order,
respectively. If we further assume all $B_{c,i}$'s are distinct and all
$\mu_i$'s have the same sign, the scattering length expression can be
factorized as:

\begin{eqnarray}
  \frac{a}{a_{\mathrm{bg}}}=\prod^N_{i=1}\frac{B-B^*_i}{B-B_{0,i}}. \label{a_form}
\end{eqnarray}

\noindent Here we define $B_{0,i}$ as the $i$-th lowest pole of
$a/a_{\mathrm{bg}}$, and $B^*_i$ the $i$-th lowest zero. This form is more
convenient for experimental research since zeros and poles of the scattering
length can be identified experimentally. Accordingly, the width of the $i$-th
Feshbach resonance can be defined as $\Delta_i = B^*_i - B_{0,i}$.

\section{Binding energy measurements and determination of the scattering length}

\subsection{Measurements of the binding energy}


We measure the  binding energies of ultracold Cs$_2$ Feshbach molecules in a
magnetic field range  where a weakly bound $s$-wave molecular state overlaps
with a $d$-wave and a $g$-wave state \cite{Mark2007}. The latter two states are
connected to the $d$-wave and $g$-wave Feshbach resonances located at about
$48$~G and $53.5$~G, respectively \cite{Chin2004}. The molecular spectroscopy
is performed by using a weak oscillating magnetic field. This technique,
previously applied to a $^{85}$Rb atomic sample \cite{Thompson2005} and to a
$^{85}$Rb-$^{87}$Rb mixture \cite{Papp2006}, has the advantage to determine
binding energies of weakly bound molecules to within a few percent. The
oscillating magnetic field stimulates a transition between an atom pair and a
molecule for a modulation frequency resonant to $E_b/ h $ \cite{Thompson2005,
Hanna2007}, where $h$ is the Planck constant.

Our molecular spectroscopy is performed using ultracold Cs atoms in an optical
trap. A detailed description of our experimental procedure can be found in
Ref.\,\cite{Rychtarik2004}. After several pre-cooling stages, the atoms are
fully polarized in their absolute ground state $|F\!=\!3, m_{F}\!=\!3\rangle$.
We evaporatively cool the atoms in the optical trap at a magnetic field of 27~G
\cite{surfacetrap}, where the two-body scattering length is large enough
($a\approx$ 450$a_0$, with $a_0$ the Bohr radius) to provide fast
thermalization. During the evaporation sequence, an extra magnetic field
gradient is applied to levitate the atoms against the gravity
\cite{Chin2005,Weber2003}. We typically obtain a nearly-degenerate sample of
8000 Cs atoms at $T= 100$~nK. Typical values for the density and phase space
density are $1.2 \times 10^{12}$ cm$^{-3}$ and 0.2, respectively.
\begin{figure}
\includegraphics[width=3.2 in]{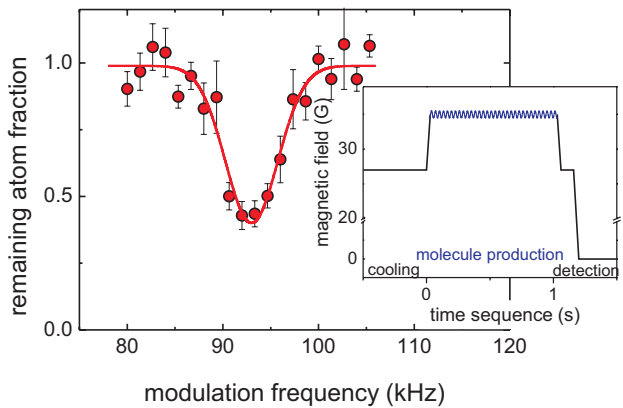}
\caption{(color online). Example of a loss resonance at 48.2 G. The losses are
due to the coupling between atom pairs and $d$-wave molecules. The solid line
is a gaussian fit to the data. We find the resonant frequency to be
$\nu=93.0(3)$~kHz and the width 5.5(6) kHz. The inset shows the typical
magnetic field sequence used to associate Cs$_2$ molecules with an oscillating
magnetic field.} \label{fig:loss}
\end{figure}

We measure the molecular binding energy by magnetic field modulation
\cite{Thompson2005,Hanna2007}. The inset of
Fig.\,\ref{fig:loss} shows the magnetic-field sequence. After ramping the
magnetic field within 3~ms to a desired value in the range of interest, we add
for typically 1~s a small sinusoidal magnetic field $B(t)=A\sin(2\pi\nu t)$,
where the modulation amplitude $A$ typically ranges from 100 to 600~mG. We then
ramp back the magnetic field to initial value and wait 100~ms, during which the
associated molecules decay by inelastic collisions with the atoms. The magnetic
field is then rapidly turned off to image the atoms with standard absorption
imaging technique. If the modulation frequency $\nu$ is close to the binding
energy value, $E_{\rm b}\simeq h\nu$, the oscillating magnetic field couples  atom
pairs to molecules \cite{Hanna2007}. The atoms are converted into molecules,
and the resulting association is detected as resonant loss in the atom number.
A typical loss signal is shown in Fig.\,\ref{fig:loss}. The resonant frequency
is determined to within a few percent  by fitting the data with a Gaussian
profile \cite{BR}.

\begin{figure}
\includegraphics[width=2.8in]{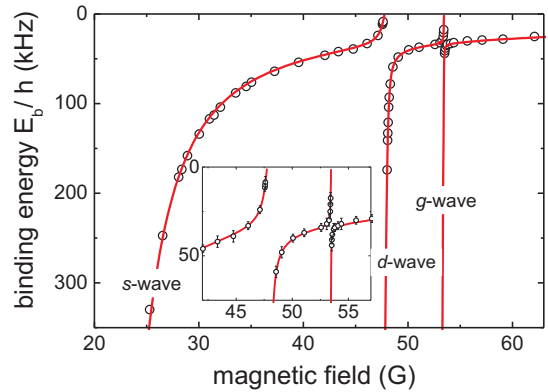}
\caption{(color online). Binding energy of cesium molecules near three Feshbach
resonances as a function of the magnetic field. Zero energy corresponds to two
Cs atoms in the absolute hyperfine ground-state sublevel $|F\!=\!3,
m_{F}\!=\!3\rangle$. The measurements are shown as open circles. The fit (solid
line) is based on Eq.\,(\ref{multi_em}), see text. The inset shows an expanded
view in the region of the two $d$- and $g$-wave narrow resonances. The error
bars refer to the statistical uncertainties. } \label{fig:csem&a}
\end{figure}

We determine the resonant loss frequency for different values of the magnetic
field. Figure \ref{fig:csem&a} shows the measured molecular spectrum (open
circles) in the magnetic field range between 25~G to 60~G. Here, three weakly
bound molecular states, namely  $s$-,  $d$-, and $g$-wave bound states, overlap
near threshold \cite{Mark2007, Hutson2008}. We usually observe losses of Cs
atoms up to 50$\%$. The largest loss rate and hence conversion efficiency is
observed when $g$-wave molecules are associated. This observation is consistent
with the results of Ref.\,\cite{Hanna2007}, which predict higher conversion
efficiencies for larger relative magnetic moments $\delta \mu$.

%

\subsection{Determination of the scattering length}

We now convert the binding energy data into scattering length. For Cs atoms,
the vdW energy  is $E_{\mathrm{vdW, Cs}}=h\times 2.7$ MHz, and the mean
scattering length $\bar a =95.7 a_B$. We apply our model in the regime $E_{\rm
b} <E_{\mathrm{vdW, Cs}}$ by fitting the binding energy data with
Eq.\,(\ref{multi_em}) for overlapping resonances. We consider three resonance
terms, which correspond to couplings to the $s$-, $d$-, and $g$-wave molecular
states. Each term contains three parameters: Feshbach coupling strength
$\Gamma_i$, bare state crossing position $B_{c,i}$ and the relative magnetic
moment $\delta\mu_i$ \cite{dmub1}. All the parameters are evaluated by the fit.
We assign the uncertainties of the parameters by a resampling method
\cite{resampling}. Second, we use the fitting parameters to determine $a$ in
the form of Eq.\,(\ref{multi_a}).
Table \ref{Table1} summarizes the main fit results for the three overlapping
resonances. We derive the zeroes and poles of the scattering length from
Eq.\,(\ref{multi_a}) and (\ref{a_form}). Our results represent the most
accurate knowledge of the positions and widths of the two experimental
important resonances near 48 and 53~G \cite{Knoop2008ooa,Danzl2008}

\begin{table}
\caption{Fitting parameters for the $s$-, $d$- and $g$-wave Feshbach
resonances, determining the scattering length in the magnetic field range of
interest; see Fig.\,\ref{fig:csem&a}. The background scattering length
$a_{\mathrm{bg}}=1875 a_0$, the mean scattering length of cesium
$\bar{a}=95.7~a_B$ and the bare $s$-wave state magnetic moment
$\delta\mu_1=2.50\mu_B$ \cite{dmub1} are set constant. Poles $B_{0,i}$ and
zeros $B^*_i$ of the scattering length are derived, see text. Uncertainties in
the parentheses are statistical. The systematic uncertainty of the magnetic
field is 10~mG.}
\begin{tabular}{cccccc}\hline
res.  & $\Gamma_i/h$(MHz) & $\delta\mu_i/\mu_B$ & $B_{c,i}$~(G) & $B_{0,i}$~(G) & $B^*_i$~(G)\\
\hline
$s$-wv. & $11.6(3)$ & 2.50  & 19.7(2) &-11.1(6) &18.1(6)\\
$d$-wv. & $0.065(3)$  & 1.15(2) & 47.962(5) &47.78(1)&47.944(5)\\
$g$-wv. & $0.0042(6)$ & 1.5(1) & 53.458(3) &53.449(3)&53.457(3)\\
\hline
\end{tabular}
\label{Table1}
\end{table}


\begin{figure}
\includegraphics[width=2.75in]{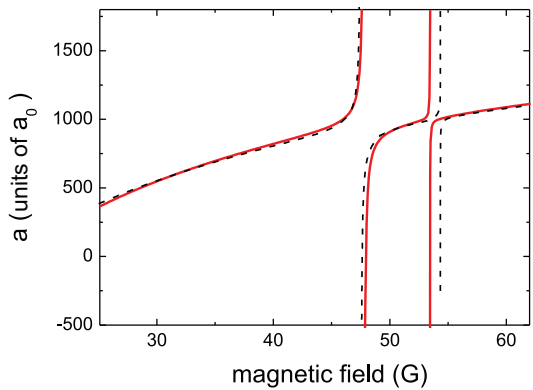}
\caption{(color online) Scattering length of $|F\!=\!3, m_{F}\!=\!3\rangle$
cesium atoms in the magnetic field range where three Feshbach resonances
overlap. The solid curve shows the result of this work while the dashed curve
represents the prediction from a previous multi-channel calculation
\cite{Chin2004}.} \label{fig:a}
\end{figure}

Figure \ref{fig:a} shows the magnetic-field dependent scattering length as derived from our model and the binding energy data. We observe a remarkable overall agreement between these results and a previous multi-channel calculation based on the knowledge of the positions of various Feshbach resonances \cite{Chin2004}. Our present method proves particularly powerful for the scattering length near the two narrow resonances. Here the previous calculations \cite{Chin2004} suffer from the large uncertainties of the positions of the resonances, which can even exceed their widths.

\section{Conclusion}

In summary, we have described a new method to precisely determine scattering lengths
from binding energy measurements of weakly bound molecular states. We employ a
simple interaction model to extract the relationship between the scattering
length and the binding energy. Using the cesium molecule binding energy data in
the range where three Feshbach resonances are present, we show that our model
provides an excellent fit to the molecular energy structure, and allows us to
precisely determine the scattering length near the resonances. The precise
knowledge of binding energy and scattering length near Feshbach resonances
provides an ideal starting point to explore few- and many-body physics by
controlling the two-body interactions. Our model is also general in the sense
that it can apply to any homonuclear or heteronuclear atomic systems near
Feshbach resonances, once molecular binding energy data are available.

\section*{Acknowledgments}

We thank E.\, Tiesinga for providing the data shown in Fig.\,\ref{fig:a} and
P.\,Julienne for discussions. We acknowledge support by the Austrian Science
Fund (FWF) within SFB 15 (project part 16). F.~F.\ is supported within the Lise
Meitner program of the FWF. C.~C.\ acknowledges the support under ARO Award
W911NF0710576 with funds from the DARPA OLE Program, the NSF-MRSEC program
under DMR-0213745 and the Packard foundation.


\begin{references}
\bibitem{surfacetrap}
The optical trap is formed by a blue-detuned evanescent laser wave (wavelength
810~nm) on top of a horizontal glass prism. In addition a red-detuned laser
beam (wavelength 1060~nm, waist 80~$\mu$m), propagating along the vertical
direction, confines the atoms in the horizontal plane.
\bibitem{BR}
We use the Breit-Rabi formula to determine the magnetic-field value from a
measurement of the resonant frequency corresponding to the $|F\!=\!3,
m_{F}\!=\!3\rangle \rightarrow |F\!=\!4,m_F\!=\!4\rangle$ atomic hyperfine
transition. Line noise limits the stability of the magnetic field to about 10
mG for typical integration times.
\end{references}
\end{document}